\documentclass[12pt,preprint]{aastex}
\begin{document}

\title{Bayesian Image Reconstruction \\
 Based on Voronoi Diagrams}

\author{G. F. Cabrera\altaffilmark{1, 2}, S. Casassus\altaffilmark{1}}
%\affil{Departamento de Astronom\'ia, Universidad de Chile, Santiago, Casilla 36-D, Chile}
\and
\author{N. Hitschfeld\altaffilmark{2}}
\email{guille@das.uchile.cl}

\altaffiltext{1}{Departamento de Astronom\'ia, Universidad de Chile,
  Santiago, Casilla 36-D, Chile} 
\altaffiltext{2}{Departamento de Ciencias de la Computaci\'on, 
  Universidad de Chile, Santiago}

%\affil{Departamento de Ciencias de la Computaci\'on, Universidad de Chile, Santiago}

\begin{abstract}
We present a Bayesian Voronoi image reconstruction technique (VIR) for
interferometric data. Bayesian analysis applied to the inverse problem
allows us to derive the \emph{a-posteriori} probability of a novel
parameterization of interferometric images. We use a variable Voronoi
diagram as our model in place of the usual fixed pixel grid.
% To describe the probability we calculate the likelihood
% and \emph{a-priori} probabilities, which need the intensity
% distributions to be discretized. We use a $\sigma_\mathrm{q}$ quantum
% for this purpose. 
A quantization of the intensity field allows us to calculate the
likelihood function and \emph{a-priori} probabilities.  The Voronoi
image is optimized including the number of polygons as free
parameters.
% We show reconstructions made over visibilities simulated using the
% Cosmic Background Imager (CBI) and compare them with a MEM
% reconstruction.
We apply our algorithm to deconvolve simulated interferometric data.
Residuals, restored images and $\chi^2$ values are used to compare our
reconstructions with fixed grid models. VIR has the advantage of
modeling the image with few parameters, obtaining a better image from
a Bayesian point of view.
%as required by Bayesian theory.
\end{abstract}


\section{Introduction}
%\subsection{Interferometric Reconstruction Basics}
% \subsection{Introduction to Image Reconstruction for Interferometric Data}

% Interferometers sample the Fourier transform of the sky intensity
% field.

% A general problem in astronomy is to obtain the sky image from the
% data. The data consists in the image convolved with the instrument
% response. 

% Astronomical interferometric data are the sum of instrumental noise
% with the convolution of the sky image and the instrumental response.
Astronomical interferometric data result from the addition of
instrumental noise to the convolution of the sky image and the
instrumental response.  Because of incomplete sampling in the $(u, v)$
plane, obtaining sky images from interferometric data is an instance
of the inverse problem, and involves reconstruction algorithms.

%\subsection{Existing Methods} 

% TESIS
%The CLEAN algorithm was created by \cite{CLEAN}. It starts with a
%direct Fourier transform of the visibilities by which it obtains the
%dirty map (DM). Then, on each iteration it subtract a dirty beam (DB)
%image centered in the maximum intensity point of the image, normalized
%by a $\gamma I_0$ value, where $I_0$ is the maximum intensity value
%and $\gamma$ is called the \emph{loop gain}. This is repeated until
%the noise level on the map is reached. The final reconstructed map
%consists of the sum of the resulting map with all the components
%removed on previously iterations but in the form of clean beams. 

The CLEAN method consists of modeling the side-lobe disturbances and
subtract them iteratively from the dirty map \citep[][]{CLEAN}. The
CLEAN method works well for low noise and simple sources. But if the
source has many complex features, or if the data is too noisy, CLEAN
will do only a few iterations returning a noisy image
\citep[][]{CLEAN}. Another shortcoming is that CLEAN involves some
ad-hoc parameters (the loop gain, stopping criteria, clean beam) that
bias the final reconstruction, in the sense that CLEAN can give many
different reconstructions for the same dataset.
% image user-dependent
% (sesgada?).

The maximum entropy method (MEM) finds the image that simultaneously
best fits the data, within the noise level, and maximizes the entropy
$S$. This is done by minimizing
\begin{equation}
  L_\mathrm{MEM} = \chi^2 - \lambda S,\label{eq:LMEM}
\end{equation}
where, for the case of interferometric data, $\chi^2$ can be
calculated as
\begin{equation}
  \chi^2 = \sum_{k=1}^{N_\mathrm{Vis}}\frac{||V_k^\mathrm{obs} -
    V_k^\mathrm{mod}||^2}{\sigma_k^2},
\end{equation}
where the sum runs over all the $N_\mathrm{Vis}$ visibilities, the
symbol $||z||$ stands for the modulus of the complex number $z$ and
$\sigma_k$ is the root mean square (rms) noise of the corresponding
visibility. $\lambda$ is a control parameter and the entropy $S$
varies for different implementations \citep[e.g.][]{Nar&Nit}. The
entropy is used as a regularizing term in a degenerate inverse
problem, when there are more free parameters than data. Different
formulations for $S$ appear in the literature. Some examples are
$\sum_i\ln(I_i)$, $\sum_iI_i\ln(I_i)$, $\sum_i\ln(p_i)$,
$\sum_ip_i\ln(p_i)$, where $I_i$ is the specific intensity value at
pixel $i$ and $p_i = I_i/\sum_iI_i$ \citep[see][and references
therein]{P&P}.

\cite{Corn&Ev} used MEM in the AIPS VM task. Their method makes some
approximations that diagonalize the Hessian matrix required to
optimize their merit function. They used an entropy of the form $S =
-\sum_iI_i\log{(I_i/ m_i)}$, where the sum extends over all the pixels
$i$, $\{I_i\}_{i=1}^n$ is the model image and $\{m_i\}_{i=1}^n$ is a
prior image. However, the neglect of the side-lobe contribution to the
Hessian may lead the optimization to local minima that still bear
instrumental artifacts. \cite{Casassus}implemented a MEM algorithm
based on the conjugate gradient method, without the use of the
Cornwell and Evans approximation.
%\cite{Casassus} have used a similar approach but without the use of
%any approximation. 
They used an entropy of the form $S = -\sum_iI_i\log{(I_i/ M)}$, where
$\{I_i\}_{i=1}^N$ is the model image and $M$ is a small intensity
value, i.e they start with a blank image prior, and $M$ is an
intensity value much smaller than the noise.
% Casassus et al. created their own MEM
% because the neglect of the side-lobes (visible in the dirty map
% surrounding the primary beam half maxima contour) contribution to the
% Hessian of the merit function in Cornwell and Evans MEM was not well
% suited to their ends.
% They did not use Cornwell and Evans MEM because their neglect of the
% side-lobes (visible in the dirty map surrounding the primary beam half
% maxima contour) contribution to the Hessian of the merit function was
% not well suited to their ends.

% The idea of introducing the entropy is to smooth the final image.

Bayesian analysis is a powerful tool for image reconstruction
techniques. In this application, our goal is to find the most probable
image by maximizing its \emph{a-posteriori} probability. For Bayesian
methods, the \emph{a-priori} and likelihood distributions are
needed. To derive the \emph{a-priori} probability the definition of an
intensity quantum is needed. This quantum represents the minimum
measurable intensity unit. The intensity in each pixel can be
interpreted as a number of quanta $I_i = \sigma_\mathrm{q}N_i$, where
$I_i$ is the intensity in pixel $i$, $\sigma_\mathrm{q}$ is the
quantum size and $N_i$ the number of quanta in pixel $i$.

\cite{P&P} used Bayesian analysis in the Pixon algorithm. They use a
variable model and maximize $P(I,M|D)$, that is, the probability of
the image $I$ and model $M$ given the data $D$. In their approach the
model used to parameterize the image is a set of Gaussians which are
used to average a pseudo-image. The pseudo-image starts as a maximum
residual likelihood reconstruction and a local Gaussian pixon is
assigned to each of its pixels. The number of pixons, and hence the
number of free parameters, is reduced in each iteration. 

\cite{W&S} have used Bayesian analysis for interferometric data, but
using a fixed pixel grid to parameterize the model image. They use
Gibbs sampling to determine the posterior density distribution.

The most typical model used in astronomy to represent the sky
brightness distribution consists of a pixel grid.  A big disadvantage
of this grid is that the number of pixels remains fixed as well as
their size. Often, uniform pixel grids involve more free parameters
than really needed to fit the data.

The purpose of this paper is to explore Bayesian reconstruction with
image models based on Voronoi tessellations in place of the usual
pixelated image. We call this new deconvolution method ``Voronoi image
reconstruction'' (VIR, hereafter). The advantage of using Voronoi
models is that it is possible to use a smaller number of free
parameters, as required by Bayesian theory. Our purpose is not optimal
CPU efficiency; we search for the optimal image and model from a
Bayesian point of view.

\begin{figure}[h]
\epsscale{.50}
\plotone{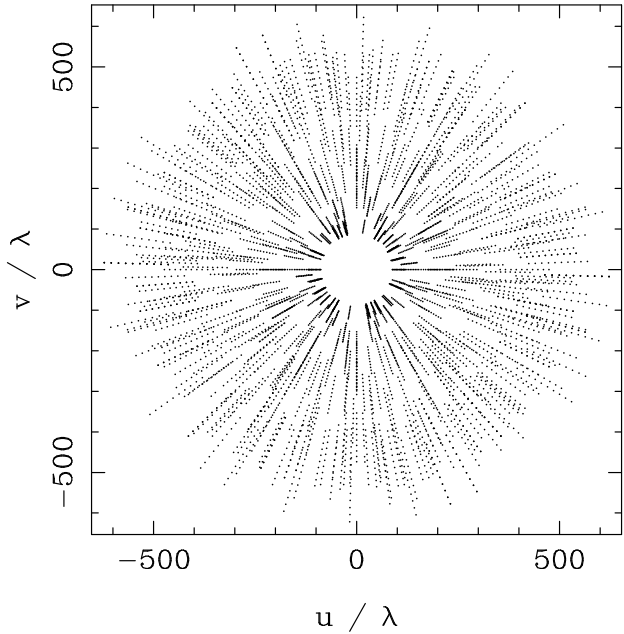}
\caption{Coverage in the $(u, v)$ plane of the CBI in the
  configuration used for our simulations.\label{fig:baselines}}
\end{figure}

We used the Cosmic Background Imager \citep[CBI,][]{pad02} to
illustrate our method. The CBI is a planar interferometer array with
13 antennas, each 0.9~m in diameter, mounted on a 6~m tracking
platform. An example of CBI baselines is shown in Figure
\ref{fig:baselines}. The radius of the hole at the center of the $(u,
v)$ plane is the reciprocal of the minimum distance between two
antennas, measured in wavelengths. The side-lobes of the CBI are
caused mainly by this central hole in the $(u, v)$ baselines.
% This central hole in the $(u, v)$ baselines causes the side-lobes of
% the CBI.

We briefly summarize the elements of Bayesian theory that determine
the probability distributions concerning our problem
(Section~\ref{sec:Bayes}).  The new model based on Voronoi
tessellations is described (Section~\ref{sec:Voronoi}), as well as
optimization issues involved in our problem
(Section~\ref{sec:Optimization}).  We discuss implementation details
such as the optimal quantum size and number of Voronoi polygons
(Section~\ref{sec:Implementation}), compare reconstructions made with
MEM and VIR (Section~\ref{sec:Results}) and finally summarize our
results (Section~\ref{sec:Conclusions}).

% The Cosmic Background Imager (CBI, \cite{pad02}) data is a particular
% case where image reconstruction is difficult. The CBI is a planar
% interferometer array with 13 antennas, each 0.9~m in diameter, mounted
% on a 6~m tracking platform, which rotates in parallactic angle to
% provide uniform $uv$-coverage. The CBI receivers operate in 10
% frequency channels, with 1~GHz bandwidth each, giving a total
% bandwidth of 26--36~GHz. It is located in Llano de Chajnantor,
% Atacama, Chile. An example of CBI baselines is shown in Figure
% \ref{fig:baselines}. The hole in the center of the $(u, v)$ plane is
% caused because of the minimum distance that two antennas can be
% arranged. This sparse distribution of the $(u, v)$ baselines causes
% the CBI to have strong side-lobes.

\section{Bayesian Theory} \label{sec:Bayes}

An image model is required to parameterize the sky brightness
distribution. The most typical model used in astronomy is a
rectangular grid of uniform pixels. That configuration of pixels is
the model $M$, and the distribution of brightness in the model is
called an image $I$.  We search for the image that represents as
accurately as possible the visibility data $D$.  The Bayesian image
reconstruction approach, using a fixed model, tries to find the image
that maximizes the probability $P(I | D, M)$, i.e. find the most
probable image given the data and the model.

Using the Bayes theorem, we obtain
\begin{equation}
  P(I | D, M) = \frac {P (D| I, M)P(I | M)}{P (D | M)}.
\end{equation}
Since the data is fixed, $P(D | M)$ is a constant in the problem when
the model is not considered as a variable. Thus, the fixed image model
optimization problem reduces to
\begin{equation}
  \max_I P(I | D, M) = \max_I P (D| I, M)P(I | M). \label{eq:P(I|D|M)}
\end{equation}
The first term, $P (D|I, M)$ is called the likelihood, and measures
 how well our data represents our image. The second term, $P (I | M)$
 is called the image prior, and gives the \emph{a-priori} probability
 of the image given the model, i.e. how probable is the image given
 only the model.

In the case of having a variable model, what we would like to find is
the image and model that maximize $P(I, M| D)$, i.e. find the most
probable image and model given the data. In this case we find
 \begin{eqnarray}
   P(I, M| D) & = & P(I|D,M)P(M|D)\nonumber\\
   & = & \frac {P (D| I, M)P(I| M)P(M|D)}{P (D | M)}\nonumber\\
   & = & \frac {P (D| I, M)P(I| M)P(M)}{P (D)}.
 \end{eqnarray}
Since the data is fixed, $P(D)$ is constant in our problem. As we
cannot privilege one model over another in the absence of image and
data, $P(M)$ is the same for all models, so it is not important for
our analysis. This way, our optimization problem reduces to
% the same one we would have with a fixed model in Eq. \ref{eq:P(I|D|M)}
 \begin{equation}
  \max_{I, M} P(I, M| D) = \max_{I, M} P (D| I, M)P(I | M).
 \end{equation}

\subsection {Probability Distributions}

Our data is a set of $N_\mathrm{Vis}$ observed visibilities
$\{V_1^\mathrm{obs}, V_2^\mathrm{obs}, \cdots,
V_{N_\mathrm{Vis}}^\mathrm{obs}\}$.  If we have a certain model $M$
and image $I$, we obtain model visibilities $\{V_k^\mathrm{mod}\}$
by simulating the interferometric observations over our image:
\begin{equation} 
V^\mathrm{mod}_k = V^\mathrm{mod}(u_k,v_k) = \int_{-\infty}^{+\infty} A(x,y)
I(x,y)\exp\left[2\pi i (u_kx+v_ky)\right]
\frac{dx\,dy}{\sqrt{1-x^2-y^2}} ~, \label{eq:vmodel}
\end{equation}
where $\{u_k, v_k\}$ are the coordinates of baseline $k$ in the $(u,
v)$ plane and $A$ is the primary beam. We thus have a set of
$N_\mathrm{Vis}$ model visibilities.  Assuming that each visibility is
independent from the others and Gaussian noise, the likelihood is
\begin{eqnarray}
  P (D|I, M) & = &
  P(\{V_k^\mathrm{obs}\}_{k=1}^{N_\mathrm{Vis}}|\{V_k^\mathrm{mod}(I, M)\}_{k=1}^{N_\mathrm{Vis}})
  = \prod_{k=1}^{N_\mathrm{Vis}} P (V_k^\mathrm{obs}|V_k^\mathrm{mod}) \nonumber\\
  & = &
  \prod_{k=1}^{N_\mathrm{Vis}}\frac{1}{2\pi\sigma_k^2}e^{-||V_k^\mathrm{obs}
  - V_k^\mathrm{mod}||^2/2\sigma_k^2}.
\end{eqnarray}
% where the symbol $||z||$ stands for the modulus of the complex number
% $z$ and $\sigma_k$ is the rms noise of the corresponding visibility.

To obtain the image prior, $P (I | M)$, we calculate the statistical
weight of a given distribution of counts \citep[as
in][]{P&P,W&S}. Consider a model consisting of $n$ cells. In the case
of a traditional image, each pixel would be a cell. There is a number
of $N$ quanta falling into these cells. These are intensity quanta of
some size $\sigma_\mathrm{q}$.
% These quanta are used to determine the number of distinguishable
% events in each cell.
In the case of a pixelated image, the intensity in each pixel $i$
would be $I_i = \sigma_\mathrm{q} N_i$, where $I_i$ is the intensity
in cell $i$. Each quantum could fall into any of the $n$ cells, so the
total number of possible configuration for the $N$ quanta will be
$n^N$. The probability of the image given the model is the probability
of a certain state $\{N_1, N_2, \cdots, N_n\}$ that represents that
image, where $N_i$ is the number of quanta in cell $i$.  Consider a
given image configuration defined by a particular distribution
$\{N_i\}$. The image distribution is not changed in the $N!$ possible
redistributions of counts between cells, provided each $N_i$ is
constant. The $\prod_i N_i!$ swaps of counts within each cell keep the
same image configuration.
% If we place all the quanta in the desired configuration, it is
% possible to swap them and still have the same configuration
% $\{N_i\}$. The number of possible swaps are $N!$. But swapping between
% elements of the same cell would leave the same state.
The model $M$ consists of the Voronoi diagram and the total number of
quanta (i.e. n, the position of the generators and N), thus the
\emph{a-priori} probability is
% reduces to $n$, thus the \emph{a-priori} probability is
\begin{equation}
P (I | M) = P (\{N_i\}|n, N) = \frac{N!}{n^N\prod_i N_i!}.
\label{eq:apriori}
\end{equation}
% \cite{P&P} and \cite{W&S} have used this same Bayesian approach for
% defining their regularizing term.

As explained above, $\sigma_\mathrm{q}$ is an intensity quantum. It is
also possible to describe the number of quanta per cell using a flux
quantum $\sigma_i^\mathrm{F}$, where $i$ is the index of the cell to
which we associate the quantum. This flux quantum can be expressed in
terms of the intensity quantum as $\sigma_i^\mathrm{F} =
\sigma_\mathrm{q}A_i$, where $A_i$ is the area of cell $i$. In this
case, the number of quanta per cell is $N_i =
F_i/\sigma_i^\mathrm{F}$, where $F_i = I_i A_i$ is the flux of cell
$i$. This leads to $N_i = I_i/\sigma_\mathrm{q}$, which is the same
expression for $N_i$ obtained using the intensity quantum
$\sigma_\mathrm{q}$. Using these cell-dependent flux quanta, the
probability of a quantum falling into each cell will be $\frac{1}{n}$
for every cell, leaving the \emph{a-priori} probability the same as
Eq. \ref{eq:apriori}.

\subsection {MEM and Natural Entropy}

% As said before, MEM tries to obtain an image that adjusts to the data
% within the noise level while maximizes the entropy. This is done by
% minimizing $L = \chi^2 - \lambda S$.

In Bayesian theory, for a fixed model, the image $I$ can be found by
optimizing the \emph{a-posteriori} probability:
\begin{eqnarray}
  \max_I P (I | D, M) & = & \min_I(-\ln{P (D| I, M)P(I | M)})\nonumber\\
  & = & \min_I\sum_{k=1}^{N_\mathrm{Vis}}\frac{||V_k^\mathrm{obs} -
  V_k^\mathrm{mod}||^2}{2\sigma_k^2} - \ln\bigg(\frac{N!}{n^N\prod_i
  N_i!}\bigg)\nonumber\\
  & = & \min_I\frac{1}{2}\chi^2 - S,\label{eq:funcL}
\end{eqnarray}
where we have defined the natural entropy $S =
\ln\bigg(\frac{N!}{n^N\prod_i N_i!}\bigg)$. \cite{W&S} call the term
$\ln{(N!/\prod_i N_i!)}$ the multiplicity prior. In the limit of large
$N_i$,
\begin{eqnarray}
S 
% TESIS
% & = & \ln{N!} - N\ln{n} - \sum_i\ln{N_i!}\nonumber\\
% & \simeq & N\ln{N} - N -N\ln{n} - \sum_i(N_i\ln{N_i} - N_i)\nonumber\\
% & = & N\ln{N} - N\ln{n} - \sum_i N_i\ln{N_i}\nonumber\\
& \simeq & N\ln\frac{N}{n} - \sum_i N_i\ln{N_i},\label{eq:SStirling}
\end{eqnarray}
and it can be seen that the Bayesian method is very similar to MEM in
the sense that we are adjusting the image to the data while maximizing
an entropy of the form of Eq. \ref{eq:SStirling}.
% \cite{W&S} made a similar analysis for the entropic term, where they
% call the term $\ln{(N!/\prod_i N_i!)}$ the multiplicity prior.
% We used the exact version of the natural entropy, but did this entropy
% analysis to compare the MEM algorithm and the Bayesian method.
VIR uses the natural entropy as a regularizing term.

% Write about MEM, the election of the entropy function and the Bayesian entropy.

\section {A New Image Model based on Voronoi Diagrams} \label{sec:Voronoi}

\begin{figure}[h]
\epsscale{.50}
\plotone{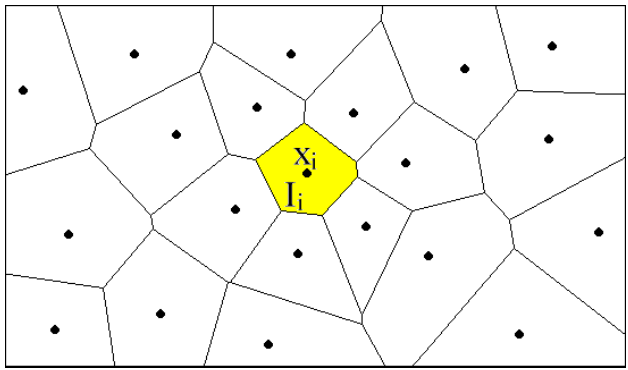}
\caption{Example of Voronoi diagram.\label{fig:Voronoi}}
\end{figure}

A Voronoi diagram is a division of the Euclidian plane into $n$
regions $\mathcal{V}_i$ defined by $n$ points $\vec{x_i}$ (called
sites or generators) such that every coordinate $\vec{x}$ in the space
belongs to $\mathcal{V}_i$ if and only if $||\vec{x} - \vec{x_i}|| <
||\vec{x} - \vec{x_j}||\ \forall\ j\neq i$. The result of the above
definition is a set of polygons defined by the generators. Figure
\ref{fig:Voronoi} shows an example of a Voronoi diagram. For further
details on Voronoi diagrams see \cite{Voronoi}.

We propose a 2D Voronoi diagram in place of the usual pixelated,
uniform grid, image as our model. We associate an intensity $I_i$ to
each of these polygons. The advantage of using a Voronoi diagram is
that we can use just as many cells (i.e. free parameters) as the data
requires. Our optimization parameters will be the position of each
generator $\vec{x_i} = (x_i, y_i)$, and the intensity at each cell,
$I_i$.

With our new model $M$ consisting of $n$ generators ($3\times n$
parameters, $x_i, y_i$ and $I_i$ for each generator), we can vary the
number of free parameters as required by the optimization problem. We
can see in equation (\ref{eq:funcL}) that the
% QUANTUM importance of the 
entropy $S$ increases as the number of cell $n$ decreases.

\section{Optimization} \label{sec:Optimization}

The optimization problem can be seen as a maximization of the
\emph{a-posteriori} probability $\max_{I, M} P (I, M | D)$, or
equivalently as a minimization of the more convenient merit function
$L = \frac{1}{2}\chi^2 - S$.
%There are different approaches for optimizing functions. 
The conjugate gradient method (CG) is often used for minimization
problems where derivatives can be easily calculated. Though it is
usually fast in convergence, CG has the problem of converging on local
minima depending on the initial condition.
% We also tried the use of
% Monte Carlo methods, such as Metropolis-Hasting, Gibbs sampling and
% some efficient Monte Carlo methods, and Genetic algorithms with no
% satisfactory results.  
The use of other optimization algorithms is postponed to future work.

% TESIS
% Monte Carlo methods (MM) and genetic algorithms (GAs) are maximization
% algorithms that try to find the global maximum of the merit
% function. MM are used for generating samples that follow a certain
% density distribution, while GAs are heuristics based on evolution of
% species that find the vector of parameters that maximize certain
% function. GA doesn't have a proof of why they work, but they often do.

%Write about the different approaches for the optimization function
%(maximization of the probability or minimization of the object
%function).

% TESIS
% \subsection{Conjugate Gradient}

The CG method searches parameters space using the gradient of the
function to be minimized. The derivatives of this function are
\begin{eqnarray}
  \frac{\partial L}{\partial x} & = & \frac{1}{2}\frac{\partial \chi^2
  }{\partial x} - \frac{\partial S}{\partial x},\\ 
  \frac{\partial \chi^2}{\partial x} & = & 2\sum_{k=1}^{N_\mathrm{Vis}}
  \frac{1}{\sigma_k^2} \mathrm{Re}\left((V_k^{\mathrm{mod}}
  - V_k^{\mathrm{obs}})^* \frac{\partial
  V_k^{\mathrm{mod}}}{\partial x}\right) \label{dLdx},
\end{eqnarray}
where $x$ is any of the optimization parameters ($x_i$, $y_i$ or
$I_i$). The derivatives of the visibilities with respect to the
position $\vec{x}_i = (x_i, y_i)$ of the $i$ generator are
\begin{eqnarray}
  \frac{\partial V_k^{\mathrm{mod}}}{\partial x_i} & = & 
  \sum_{j\in J_i} \bigg[(I_i - I_j)
    \sum_{l|\mathrm{pixel~}l\epsilon a_{ij}}A_l\Delta t_l
    (M_xt_l + b_x)e^{(t_lc_2 + s_0c_1)}\bigg],\\ 
  \frac{\partial V_k^{\mathrm{mod}}}{\partial y_i} & = & 
  \sum_{j\in J_i} \bigg[(I_i - I_j)
    \sum_{l|\mathrm{pixel~} l\epsilon a_{ij}}A_l\Delta t_l
    (M_yt_l + b_y)e^{(t_lc_2 + s_0c_1)}\bigg],
\end{eqnarray}
where $I_i$ is the intensity in cell $i$, $J_i$ is a set of the
indices of the polygons adjacent to $\mathcal{V}_i$, $a_{ij}$ is the
edge which divides polygons $\mathcal{V}_i$ and $\mathcal{V}_j$, $l$
sums over the pixels which intersect $a_{ij}$, $A$ is the CBI primary
beam. For further details see Sec.~\ref{ap:derivatives}.

The derivative of the visibilities with respect to the intensity of
each cell $I_i$ is
\begin{equation}
  \frac{\partial V_k^{\mathrm{mod}}}{\partial I_i} =
  \frac{\sin{(\pi u_k\Delta x)}\sin{(\pi v_k\Delta y)}}{\pi^2u_kv_k}
  \sum_{\textrm{\scriptsize{pixels }}l\epsilon \mathcal{V}_i}A_l
  e^{2\pi i(u_kx_l+v_ky_l)} \label{eq:dVdI},
\end{equation}
where $\vec{k}_k = (u_k, v_k)$ is the baseline corresponding to the
pair of antennas $k$, $\Delta x$ and $\Delta y$ are the pixel width
and height, and the sum extends over all the pixels inside
$\mathcal{V}_i$.

The entropy only depends of the intensities $I_i$, so $\frac{\partial
S}{\partial x_i} = \frac{\partial S}{\partial y_i} = 0$, then (see
Sec.~\ref{ap:derivativesdS})
\begin{equation} 
  \frac{\partial S}{\partial I_i} = \frac{1}{\sigma_\mathrm{q}}
  (\sum_{k=1}^{N}\frac{1}{k} - \ln{n} - \sum_{k=1}^{N_i}\frac{1}{k}).
\end{equation}

\section{VIR Design and Implementation} \label{sec:Implementation}

We have designed, and implemented in c++, VIR with 6 modules which include
algorithms for:
\begin{itemize}
  \item the generation of the Voronoi diagram
  \item calculation of model visibilities
  \item calculation of the merit function $L$ to be optimized as well
    as its derivatives
  \item fitting a Voronoi diagram to an image
  \item the CG method
  \item the optimization of the number of polygons
\end{itemize}
%We have implemented them by using object oriented programming in c++.

VIR uses the CG method from \cite{NumericalRecipes} and searches for
the position and intensities of the Voronoi polygons, ${x_i, y_i,
I_i}$, that minimize our merit function $L$. The CG method modifies
the intensities and also moves the positions of the Voronoi
generators. This causes the shape of the Voronoi polygons to change as
well.
% for optimizing the
% position and intensity of each Voronoi cell. 
% We decided to use object oriented programming and programmed a set of
% classes in C++.
A general problem with CG is that it usually converges on local
minima. For VIR in particular, though Voronoi polygons intensities
adjust quite fine, the positions of the generators are difficult to
modify substantially. The VIR parameter space is smooth enough in
intensity space to converge to a good solution. But the parameter
space in cell generator positions is very structured, and CG is
quickly stuck on local minima.

Due to the fact that CG easily falls into local minima, we needed a
good approximation for the initial Voronoi diagram. For this purpose
we used a pixelated version of the Bayesian algorithm, where the model
was a uniform grid.
% QUANTUM
% But, as explained below, the importance of the natural entropy term
% increases for many free parameters, which causes plainer images than
% expected. 
We decided to do a pure $\chi^2$ (maximum likelihood, ML)
reconstruction and use the fifth CG iteration as our starting
image. We chose this particular iteration because on inspection the
modeled images were still smooth.
%, this iteration picks up smooth intensity fields. 
Pure $\chi^2$ reaches convergence with noisy images, where the
true image is unrecognizable.
%by visual inspection of its residuals. 
We then fitted a Voronoi diagram to the image (see
Sec.~\ref{ap:fitting}) and ran CG using the positions and intensities
of the generators as our free parameters, which led to our final
reconstruction. Truncation to a level of $10^{-5}$ quanta was used to
enforce positivity.

% \subsection{Quantum Size and Number of Voronoi Generators}

\begin{figure}[h]
%\epsscale{1.50}
\plotone{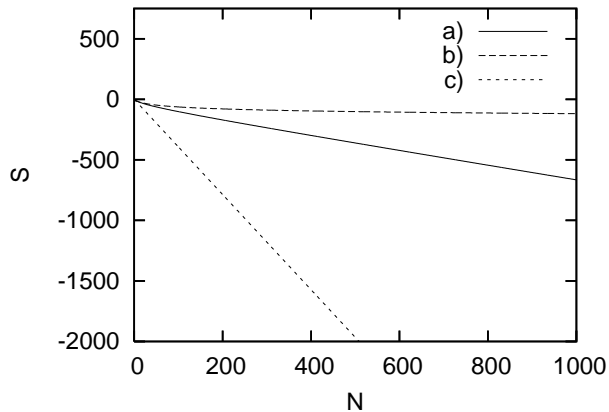}
\caption{Entropy values for different $N$, $n = 51$ and keeping
  $\{I_i\}$ fixed. This is achieved by varying
  $\sigma_\mathrm{q}$. (a) VIR reconstruction intensities. (b) Uniform
  intensities distribution, $N_i = \frac{N}{n}$ $\forall$ $i$. (c)
  Only one cell has all the quanta. $N_i = N$, $N_j =0$ $\forall$
  $j\neq i$.
  \label{fig:SvsN}}
\end{figure}

\begin{figure}[h!]
  \epsscale{1}
  \plotone{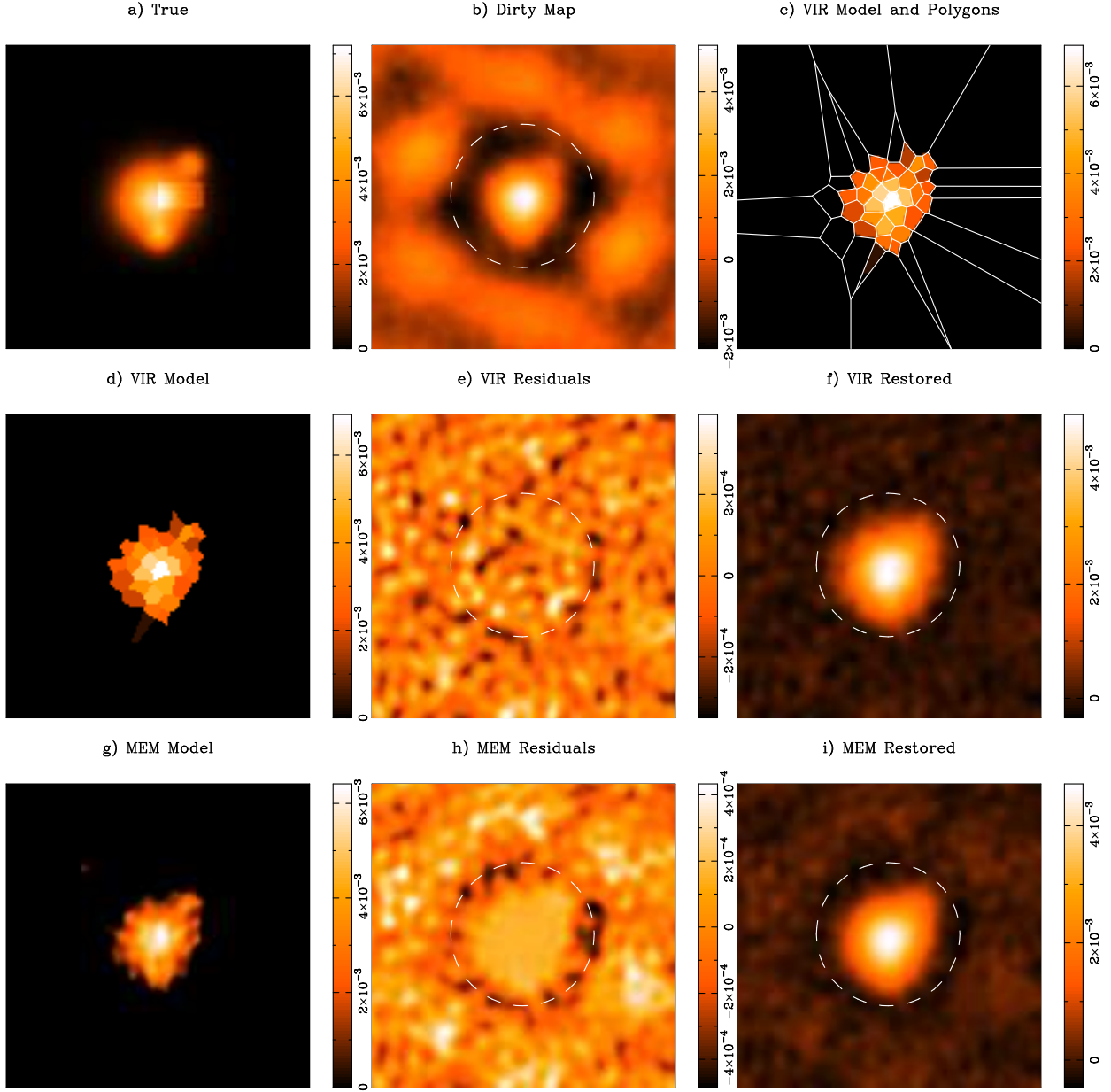}
  \caption{Comparison of MEM and VIR reconstruction techniques for a
    SNR of $\sim52$. (a) The true image. (b) Dirty map. (c) VIR
    reconstruction with its polygons drawn. (d) VIR
    reconstruction. (e) Dirty map of the VIR reconstruction
    residuals. (f) Restored image for the VIR model. (g) MEM
    reconstruction. (h) Dirty map of the MEM reconstruction
    residuals. (i) Restored image for the MEM model.
    \label{fig:reconstruccion}}
\end{figure}

An important issue to consider is the size of the quantum
$\sigma_\mathrm{q}$. \cite{W&S} treat $\sigma_\mathrm{q}$ as a free
parameter. But, as we now explain, $\sigma_\mathrm{q}$ was held
constant in this implementation of VIR. We treat the number of quanta
per cell as a continuous variable in order to use the CG method.
Entropy is maximized at $\sigma_\mathrm{q} = \infty$, where, for a
given configuration of intensities $\{I_i\}$, $N = 0$ and $S = 0$. For
every other value of $N$, the entropy will be negative. This means
that even for large $\sigma_\mathrm{q}$, the intensities $I_i =
\sigma_\mathrm{q}N_i$ can have reasonable values (using small
$N_i$). Figure \ref{fig:SvsN} shows $S$ as a function of $N$ for $51$
Voronoi generators and 3 different intensity distributions using the
model tessellation of Figure \ref{fig:reconstruccion}c. We considered:
1- the VIR intensities of Figure \ref{fig:reconstruccion}c, 2- a
uniform intensity distribution image ($N_i = \frac{N}{n}$ $\forall$
$i$), 3- a spike where all $N$ are only in one cell ($N_i = N$, $N_j
=0$ $\forall$ $j\neq i$). The curves of Figure \ref{fig:SvsN} are
obtained by keeping the intensities fixed and modifying
$\sigma_\mathrm{q}$ in order to obtain different $N$. It can be seen
on Figure~\ref{fig:SvsN} that the entropy is maximized at $N = 0$,
independently of the intensities $\{I_i\}$ of the model, where the
optimal value of $\sigma_\mathrm{q} = \infty$ is achieved if the
number of quanta per cell is treated as a continuous variable. If the
number of quanta per cell were discrete variables, as in \cite{W&S},
the choice of a big $\sigma_\mathrm{q}$ would admit only zero values
for every cell. Otherwise, if one or more quanta fell in a given cell,
the intensity of that cell would diverge as $\sigma_\mathrm{q}$ for
arbitrarily large $\sigma_\mathrm{q}$, causing a big $\chi^2$
value. Therefore, in our continuous optimization the intensity quantum
must be determined a-priori.

%QUANTUM
%The importance of the entropy diminishes for small $N$. Figure
%\ref{fig:SvsN} shows $S$ as a function of $N$ for $57$ generators
%Voronoi diagrams with 3 different intensity distributions using the
%same model tessellation of Figure \ref{fig:reconstruccion}a. We
%considered: 1- the VIR intensities of Figure
%\ref{fig:reconstruccion}a, 2- a uniform intensity distribution image
%($N_i = \frac{N}{n}$ $\forall$ $i$), 3- a spike where all $N$ are only
%in one cell ($N_i = N$, $N_j =0$ $\forall$ $j\neq i$). The curves of
%Figure \ref{fig:SvsN} are obtained by keeping the intensities fixed
%and modifying $\sigma_\mathrm{q}$ in order to obtain different $N$.
%The diminution of $||S||$ when $N$ diminishes means that for larger
%$\sigma_\mathrm{q}$, $\chi^2$ becomes more important than the entropy,
%resulting in noisy optimal images. We reach this conclusion because
%%This is caused by the fact that 
%we are optimizing our $N_i$ as continuous variables, which means that
%even for large $\sigma_\mathrm{q}$, the intensities $I_i =
%\sigma_\mathrm{q}N_i$ will have reasonable values (using small
%$N_i$). If the number of quanta per cell were discrete variables, as
%in \cite{W&S}, the choice of a big $\sigma_\mathrm{q}$ would admit
%only zero values for every cell. Otherwise, if one or more quanta fell
%in a given cell, the intensity of that cell would diverge as
%$\sigma_\mathrm{q}$ for arbitrarily large $\sigma_\mathrm{q}$, causing
%a big $\chi^2$ value.
%
%Therefore, in our continuous optimization the intensity quantum must
%be determined a-priori. 
In the Bayesian description of the entropy we count events that fall
in each cell. It seems reasonable to take the noise level as the
minimum value of intensity we can distinguish. So, $\sigma_\mathrm{q}$
should approximate the estimated thermal noise in the naturally
weighted dirty map. The definition of the weighted dirty map
\citep[e.g.][]{Briggs} is
\begin{eqnarray}
  I^\mathrm{D} (x, y) \equiv
  \int_{-\infty}^{\infty}\int_{-\infty}^{\infty}W(u, v)V(u, v)e^{-2\pi
  i(ux + vy)}dudv,\\
  W(u, v) = \frac{1}{\sum_kw_k} \sum_kw_k\delta(u-u_k, v-v_k),
\end{eqnarray}
where the sums extend over all visibilities, $w_k$ are the weights
given to visibility $k$ and $\delta$ is the two-dimensional Dirac
delta function. Propagating the thermal noise, we get for the standard
deviation of the dirty map
\begin{equation}
\sigma_\mathrm{rms}^\mathrm{D} =
\sqrt{\frac{\sum_kw_k^2\sigma_k^2}{(\sum_kw_k)^2}}, \label{eq:sigmaD}
\end{equation}
where $\sigma_k$ are the visibilities standard deviations. To take
into account model pixels correlated by the interferometer beam, we
should multiply the previous expression by $\sqrt{N_\mathrm{beam}}$,
where $N_\mathrm{beam}$ is the number of pixels inside a beam
pattern. This leads to
\begin{equation}
  \sigma_\mathrm{rms} = \sqrt{\frac{\sum_kw_k^2\sigma_k^2}
  {(\sum_kw_k)^2}}\sqrt{N_\mathrm{beam}}.
\end{equation}
For natural weights, $\sigma_k^2 = \frac{1}{w_k}$,
\begin{equation}
  \sigma_\mathrm{rms} = \sqrt{\frac{N_\mathrm{beam}}{\sum_kw_k}} =
  \sqrt{\frac{N_\mathrm{beam}}{\sum_k\frac{1}{\sigma_k^2}}}.
\end{equation}
We calculated the noise with natural weighting, $w_k =
\frac{1}{\sigma_k^2}$, because this is the weight we give to each
individual visibility data in the optimization of the merit function.

\begin{figure}[h]
  \epsscale{.50}
\plotone{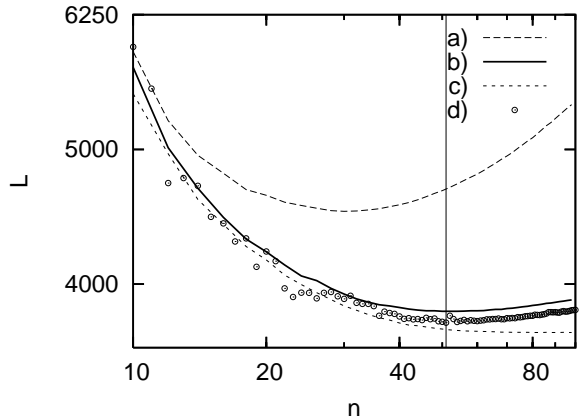}
\caption{The merit function $L$ for different $\sigma_\mathrm{q}$ and
  number of polygons $n$. The lines are averages taken over 100
  different realizations of noise for each $n$. (a) Reconstructions
  made using $\sigma_\mathrm{q} =
  \frac{1}{10}\sigma_\mathrm{rms}$. (b) Reconstructions made using
  $\sigma_\mathrm{q} = \sigma_\mathrm{rms}$. (c) Reconstructions made
  using $\sigma_\mathrm{q} = 10\sigma_\mathrm{rms}$. (d) $L$ as a
  function of $n$ for a practical application of VIR to the simulated
  visibilities used in the reconstructions of
  Figure~\ref{fig:reconstruccion}. In this practical application, the
  minimum $L$ was found at $n = 51$, and is indicated by a vertical
  line.
  %(d) Dots represent the reconstruction of our simulation for
  %different $n$. A vertical line is drawn at $n = 51$, where the
  %minimum was found. 
  \label{fig:Lvsn}}
\end{figure}

Once we have the value of $\sigma_\mathrm{q}$ we search for the
optimal number of cells $n$. In Figure \ref{fig:Lvsn} we plot the
optimal merit function for different $n$ and
$\sigma_\mathrm{q}$. These reconstructions were made over a simulation
of CBI observations on a mock sky image (Figure
\ref{fig:reconstruccion}a). We averaged over 100 reconstructions with
different realizations of Gaussian noise. The average curves shown in
Figure \ref{fig:Lvsn}, start with $n = 10$ and end with $n = 100$ for
even $n$. One single reconstruction for all $n$ took about two hours
using an AMD Athlon64 XP3000 processor with 1GB of DDR RAM at 333 MHz,
so the 300 reconstructions
% using the three different values of $\sigma_\mathrm{q}$ 
took about $25$ CPU days, but we distributed the work in six
computers, so it took about $5$ real days in total. It can be seen
that for a signal to noise ratio (SNR) of $\sim 52$, on average, the
optimal number of polygons $n$ is between 50 and 55. When
$\sigma_\mathrm{q}$ is diminished to $\frac{1}{10}\sigma_\mathrm{q}$,
on average, the optimal merit function is found at $n$ close to
$30$. For $\sigma_\mathrm{q} = 10\sigma_\mathrm{q}$, the optimal $n$
is found between $80$ and $90$.  It can be seen that as we increase
the value of $\sigma_\mathrm{q}$ we reach lower values for our
function, as discussed above. Furthermore, the optimal number of
polygons increases.
%This is because as $\sigma_\mathrm{q}$
%increases, $\chi^2$ becomes more important than the entropy giving a
%reconstruction similar to a ML, and so more parameters are needed to
%model the noise.

%\subsection{Number of Voronoi Generators}
%Optimal number of polygons (still running, I had a few errors).

\section{Example Reconstruction} \label{sec:Results}
\subsection{Mock Dataset}
% We test the VIR algorithm over CBI simulations. 
% As said before, the CBI strong side-lobes make image reconstruction a
% difficult task.
The mock sky image we used for simulations is a $256\times 256$ image
consisting of three Gaussians and a rectangle. Figure
\ref{fig:reconstruccion}a shows this image on a $128\times128$ pixel
field. Pixels are $0.75'\times0.75'$, while the CBI's primary beam is
of $45'$ FWHM (60 pixels), so most of the emission lies under the
beam. We simulated a CBI observation of 3620 visibilities over this
image and added Gaussian noise to the visibilities in order to reach a
SNR of $\sim52$. This SNR was calculated by taking the maximum
intensity from the dirty map using natural weights, and using the
noise $\sigma_\mathrm{rms}^\mathrm{D}$ (see
Eq. \ref{eq:sigmaD}). Simulation of the CBI observations is performed
with the MockCBI program (Pearson 2000, private communication), which
calculates the visibilities $V(u,v)$ on the input images $I(x,y)$ with
the same $uv$ sampling as a reference visibility dataset
(Eq.~\ref{eq:vmodel}). Thus MockCBI creates the visibility dataset
that would have been obtained had the sky emission followed the true
image. Figure \ref{fig:reconstruccion}b shows the dirty map calculated
over these simulated visibilities using the DIFMAP package
\citep[][]{she97}. The CBI's primary beam is drawn as a dashed
circle. The secondary side-lobes due to the central discontinuity in
$u$-$v$ coverage can be distinguished in Figure
\ref{fig:reconstruccion}b at a level comparable to the true emission.

\subsection{MEM Reconstruction}
The VIR method was compared with the MEM algorithm described in
\cite{Casassus}. To fit the model image to the observed visibilities,
MEM calculates the model visibilities required by its merit function
$L_\mathrm{MEM}$.
% MEM fits model visibilities, calculated on a model
% image, to the observed visibilities. 
The model visibilities are those obtained by a simulation of CBI
observations had the sky followed the model image
% (``CBI-simulated visibilities'' hereafter)
.  The free-parameters of our MEM model are the pixels in the model
$64\times64$ image. The model functional we minimize is
$L_\mathrm{MEM} = \chi^2 - \lambda S$, with the entropy $S= - \sum_i
I_i \log I_i/M$, where $M$ is a default pixel value well below the
noise level, and $\{I_i\}_{i=1}^{N}$ is the model image. We started
with the fifth iteration of a pure $\chi^2$ reconstruction ($\lambda =
0$) as initial condition for the CG minimization. This is the same ML
initial condition used in our VIR method. Figure
\ref{fig:reconstruccion}g shows the reconstructed image using $\lambda
= \frac{100}{\sigma_\mathrm{rms}}$ and $M =
10^{-2}\sigma_\mathrm{rms}$ inset on a larger $128\times128$ image
\footnote{We choose to display the sky images in a larger field than
the domain of free parameters; larger fields are required to highlight
secondary side-lobes}.

\subsection{VIR Reconstruction}
The MEM algorithm described above requires the prior assignment of the
$\lambda$ and $M$ parameters as well as the entropy formula. In
contrast, our VIR algorithm is free from such arbitrary parameters
(provided the optimal $\sigma_\mathrm{q}$ is indeed equal to
$\sigma_\mathrm{rms}$). For our VIR method, we only need to find the
number of polygons to be used. In order to find the optimal number of
polygons we reconstructed with different numbers of generators in a
range covering each natural number from $n = 6$ to $n = 100$. We found
a minimum at $n = 51$. Figure \ref{fig:Lvsn} summarizes this
search. The whole search for a particular realization of noise took
about 10 hours on the AMD Athlon64 XP3000 processor with 1GB of DDR
RAM at 333 MHz. The VIR reconstruction using 51 polygons is shown in
Figure \ref{fig:reconstruccion}c, where the Voronoi cells have also
been drawn. Figure \ref{fig:reconstruccion}d shows the same model but
without drawing the Voronoi mesh.

\subsection{Results}
% MODEL
The quality of each reconstruction can be assessed by visual
inspection, comparing the VIR and MEM model images with the true
image.  The MEM model looks similar to the true image but is
noisy. The density of Voronoi generators in the VIR model is greater
where there is more emission in the true image, approximating the true
image with only a few polygons.  We calculated $\chi^2_\mathrm{im} =
\sum_i(I_i^\mathrm{mod} - I_i^\mathrm{true})^2$, where
$I_i^\mathrm{mod}$ is the intensity at pixel $i$ of the model image
(MEM or VIR), $I_i^\mathrm{true}$ is the intensity at pixel $i$ of the
true image, and the sum extends over all pixels in the
images. $\chi^2_\mathrm{im}$ gives a measure of how well the model
fits the true image. It can be seen in Table \ref{table:funciones}
that the VIR reconstruction has a better $\chi^2_\mathrm{im}$ than
MEM, showing that the VIR model is closer to the true image than the
MEM model.

\begin{deluxetable} {crrrrr}
\tabletypesize{\small} \tablecaption {Comparison
between MEM and VIR reconstructions.\label{table:funciones}}
\tablewidth{0pt} 
\tablehead{ & \colhead{$\chi^2$} & \colhead{$\frac{\chi^2}{n_\mathrm{data}}$} 
  & \colhead{$L$} & \colhead{$\chi^2_\mathrm{im}$}}
\startdata
MEM & 7354.85 & 1.016 & 12192.6 & 0.001608 \\
VIR & 7221.04 & 0.997 & 3753.28 & 0.001396 \\
\enddata
\end{deluxetable}

% RESIDUALS
Figures \ref{fig:reconstruccion}e and \ref{fig:reconstruccion}h show
the VIR and MEM models residuals. Residual images are the dirty map of
the residuals of the visibilities, calculated over the optimal model
visibilities. It can be noted on Figure \ref{fig:reconstruccion}e that
the VIR residuals are very good, showing only noise. On the other
hand, in the MEM residuals (Figure \ref{fig:reconstruccion}h) the
object shape can clearly be distinguished as well as the CBI's
side-lobes. The object seems to be more compact in the model than in
its MEM residuals; as expected these residuals are convolved with the
synthetic beam.

% RESTORED
Restored images
%(convolution of the reconstruction with the CBI beam plus the residuals) 
are shown in Figures \ref{fig:reconstruccion}f and
\ref{fig:reconstruccion}i. These images are obtained by convolving the
models with a Gaussian point spread function (PSF) given by DIFMAP and
adding the dirty map of the residuals visibilities.
% The restored images show similar results. 
On Figures \ref{fig:reconstruccion}f and \ref{fig:reconstruccion}i it
can be assessed that VIR produces improved restored images relative to
MEM. The VIR restored image is similar to that expected given the
instrumental noise: it approximates the true image convolved with a
Gaussian PSF plus a uniform noise level. In the MEM restored image, on
the other hand, the CBI side-lobes can still be distinguished.

% STATS
The number of optimization parameters in MEM are $64\times64 = 4096$,
while the VIR method has only $51$ triplets (cell's $(x, y)$ position
and intensity) i.e. $153$ free parameters. This smaller number of
parameters causes the Bayesian entropy to be greater than the
pixelated version, obtaining a smaller value for our merit function
$L$ to be minimized.

Table \ref{table:funciones} also shows
$\frac{\chi^2}{n_\mathrm{data}}$ values, where $n_\mathrm{data}$ is
the number of data points ($3620\times 2$ in our case). A good
reconstruction should have a $\frac{\chi^2}{n_\mathrm{data}}$ value
close to $1$. It can be seen that the VIR model gives a value of
$\frac{\chi^2}{n_\mathrm{data}}$ closer to $1$ than the MEM
reconstruction.
% The VIR model is more satisfactory in this $\chi^2$ sense than the MEM
% model.

% Table \ref{table:funciones} shows different parameters used to
% evaluate MEM and VIR reconstructions. We calculated two kinds of
% reduced $\chi^2$. For a large number of parameters $n$, the use of a
% reduced $\chi^2$ of the form $\chi^2 = \frac{\chi^2}{m}$ is needed,
% where $m$ is the number of data points ($3620\times 2$ in our
% case). When using a smaller $n$ the reduced $\chi^2$ is $\chi^2_\nu =
% \frac{\chi^2}{\nu}$, where $\nu = m - n$ is the number of degrees of
% freedom. In MEM models the number of free parameters $n$ can be larger
% than $m$. It can be seen in Table \ref{table:funciones} that the MEM
% reconstruction has a reduced $\chi^2$ value close to 1 only when using
% the first definition. By contrast, the VIR reconstruction has a
% reduced $\chi^2$ close to 1 for both definitions. VIR models are
% satisfactory in the reduced $\chi^2$ sense, while MEM models lack a
% statistical interpretation.

% MEM model does not give a good fit in this sense.
% Although it cannot be strictly comparable, the Bayesian $L$ function
% gives a much smaller value for VIR than for MEM (the MEM model
% optimizes $L_\mathrm{MEM}$, not $L$). This reflects the increase in
% entropy with the number of free parameters.

\section{Conclusions} \label{sec:Conclusions}
We have introduced a Bayesian Voronoi image reconstruction (VIR)
technique for interferometric data where the image is represented by a
Voronoi tessellation in place of the usual pixelated image. The
advantage of Voronoi models is that we can use a smaller number of
free parameters, as required by the Bayesian analysis of a discretized
intensity field. Our purpose is not optimal CPU efficiency; we search
for the optimal image and model from a Bayesian point of view. The
free parameters of our model are the Voronoi generators positions
$(x_i, y_i)$ and intensities $I_i$. The following points summarize our work:
\begin {itemize}
% \item We used a new model for representing the sky plane consisting of
%   a Voronoi tessellation. The free parameters of our model are the
%  Voronoi generators positions $(x_i, y_i)$ and intensities $I_i$.
\item We discretized the intensity field in order to calculate \emph{a
  priori} probabilities. We defined a quantum intensity value
  $\sigma_\mathrm{q}$ such that $I_i = \sigma_\mathrm{q} N_i$, where
  $I_i$ is the intensity at cell $i$ and $N_i$ the number of quanta in
  cell $i$.
% \item We defined our merit function as $-\ln{P (I, M | D)}$ and used a
%   conjugate gradient algorithm for optimizating it.
\item We calculated the analytical derivatives required by the
  conjugate gradient and cross checked them by finite
  differences. Because the parameter space in cell generators
  positions is very structured, the positions of the Voronoi
  generators are difficult to change. As initial condition we took a
  Voronoi tessellation adjusted to an interrupted maximum likelihood
  reconstruction.
  % (making the position of the  generators 
  % hard to move), we adjusted a Voronoi tessellation to the fifth 
  % iteration of a maximum likelihood reconstruction as initial 
  % condition.
\item We simulated a CBI observation over a true image and
  reconstructed sky images from this mock visibility dataset using MEM
  and VIR.
\item We defined the value of $\sigma_\mathrm{q}$ as the estimated
  noise of the dirty map and searched for the optimal number of
  Voronoi polygons for our example dataset.
\item We finally compared the MEM and VIR models, residuals and
  restored images. The VIR model is closer to our true image than the
  MEM model. Residuals and restored images are also better in VIR than
  in MEM. We found that VIR model visibilities give a better fit to
  the data than MEM, in the sense that $\chi^2$ is closer to its
  expected value.
  
  %We also calculated $\chi^2$ values for quantitative comparison,
  %obtaining a value closer to the number of data points for VIR.
\end{itemize}

% Analysis of the correct quantum size for the \emph{a-priori} probability was
% also made as well as its dependency in the optimal number of polygons.
%
% Our implementation consists in a conjugate gradient method which
% maximizes the \emph{a-posteriori} probability of the image. Function
% derivatives with respect to Voronoi generators and intensities are
% needed, calculated analytically and cross checked by finite
% differences.
%
% Reconstructions over simulated data on a true image using our
% algorithm and a MEM one were compared.

\acknowledgments
We are grateful to Tim Pearson for advice on FFTs and the use of
MOCKCBI. G.F.C. and S.C. acknowledge support from FONDECYT grant
1060827, and from the Chilean Center for Astrophysics FONDAP 15010003.

\appendix

\section{Derivatives}\label{ap:derivatives}

Our merit function for minimization is
\begin{eqnarray}
L & = &
\frac{1}{2}\sum_{j=1}^{N_\mathrm{Vis}}\frac{||V_j^{\mathrm{mod}}
  - V_j^{\mathrm{obs}}||^2}{\sigma_j^2}
-\ln\left(\frac{N!}{n^N \prod_{i=1}^{n}N_{i}!}\right)\\ & = &
\frac{1}{2}\chi^2 - S.
\end{eqnarray}
So, the derivative of $L$ with respect to any variable $x$ is
\begin{equation}
\frac{\partial L}{\partial x} = \frac{1}{2}\frac{\partial
\chi^2}{\partial x} - \frac{\partial S}{\partial x}
\end{equation}

\subsection {Calculation of the Derivatives of $\chi^2$}

$\chi^2$ derivatives with respect to any variable $x$ can be obtain as follows
\begin{eqnarray}
\frac{\partial}{\partial x} \frac{1}{2}\chi^2
  & = & \frac{\partial}{\partial x}
    \left(\frac{1}{2}\sum_{k=1}^{N_\mathrm{Vis}}\frac{||V_k^{\mathrm{mod}} -
    V_k^{\mathrm{obs}}||^2}{\sigma_k^2}\right)\nonumber\\
% TESIS
%   & = & \frac{1}{2}\sum_{k=1}^{N_\mathrm{Vis}} \frac{1}{\sigma_k^2}
%     \frac{\partial}{\partial x}
%     \left((V_k^{\mathrm{mod}} -
%     V_k^{\mathrm{obs}})(V_k^{\mathrm{mod}} -
%     V_k^{\mathrm{obs}})^*\right)\nonumber\\
%   & = & \frac{1}{2}\sum_{k=1}^{N_\mathrm{Vis}} \frac{1}{\sigma_k^2}
%     \frac{\partial}{\partial x}
%     \left(\mathrm{Re}(V_k^{\mathrm{mod}} - V_k^{\mathrm{obs}})^2
%     + \mathrm{Im}(V_k^{\mathrm{mod}} - V_k^{\mathrm{obs}})^2
%     \right)\nonumber\\
  & = & \sum_{k=1}^{N_\mathrm{Vis}} \frac{1}{\sigma_k^2}
    \left(\mathrm{Re}(V_k^{\mathrm{mod}} - V_k^{\mathrm{obs}})
    \mathrm{Re}\left(\frac{\partial V_k^{\mathrm{mod}}}{\partial x}\right)
    %\nonumber\\ &&
    + \mathrm{Im}(V_k^{\mathrm{mod}} - V_k^{\mathrm{obs}})
    \mathrm{Im}\left(\frac{\partial V_k^{\mathrm{mod}}}{\partial x}\right)
    \right),\nonumber\\
  && \label{eq:dLdI}
\end{eqnarray}
where its necessary to calculate the model visibilities derivatives
with respect to $x$.

\subsubsection {Calculation of $\frac{\partial V_k^{\mathrm{mod}}}{\partial I_i}$}

% As said before, $V(\vec{k}) = \int A(\vec{x}) I(\vec{x})
% e^{2\pi i\vec{k}\vec{x}}d\vec{x}$, but according to 
In our Voronoi tessellation representation of the sky image
\begin{equation} \label{eq:VisDiscreta}
V(\vec{k}) = \sum_i^{N_\mathrm{V}} I_i\int_{\mathcal{V}_i} A(\vec{x}) e^{2\pi
i\vec{k}\vec{x}}d\vec{x},
\end{equation}
where $N_\mathrm{V}$ is the number of polygons, $\mathcal{V}_i$ is
polygon $i$ and $I_i$ its intensity. We neglected the $\sqrt{1 - x^2 -
y^2}$ term which is close to $1$, but it can easily be included in
$A(\vec{x})$. After derivation and defining $f_k(\vec{x})\equiv
A(\vec{x}) e^{2\pi i\vec{k_k}\vec{x}}$ we obtain
\begin{eqnarray}
\frac{\partial V_k^{\mathrm{mod}}}{\partial I_i} & = &
\int\int_{\mathcal{V}_i} f_k(\vec{x})d^2x, \\
& = & \frac{\sin{(\pi u_k\Delta x)}\sin{(\pi v_k\Delta y)}}{\pi^2u_kv_k}
  \sum_{\textrm{\scriptsize{pixels }}l\epsilon \mathcal{V}_i}A_l
  e^{2\pi i(u_kx_l+v_ky_l)}, \\
& \simeq & \Delta x\Delta y
  \sum_{\textrm{\scriptsize{pixels }}l\epsilon \mathcal{V}_i}A_l
  e^{2\pi i(u_kx_l+v_ky_l)}
\end{eqnarray}
for small $\Delta x$ and $\Delta y$.
% This integral can be computed numerically, and its domain is only
% polygon $i$.

\subsubsection {Calculation of $\frac{\partial V_k^{\mathrm{mod}}}{\partial x_i}$ and $\frac{\partial V_k^{\mathrm{mod}}}{\partial y_i}$}

\begin{figure}[h]
%  \plotone{figuras/Voronoi_deltaB&N.ps}
  \plotone{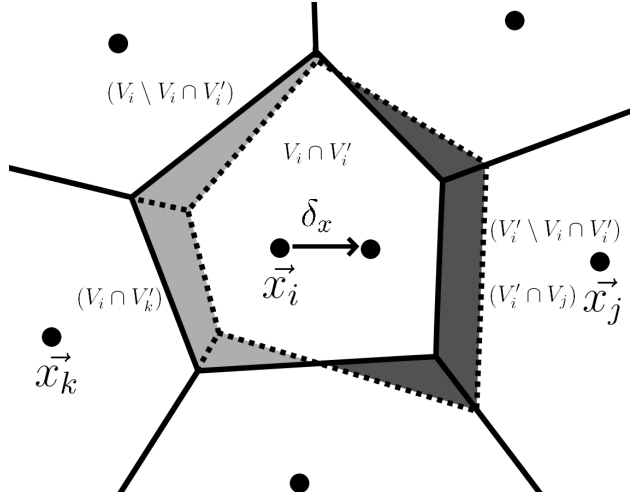}
  \caption{Voronoi tessellation before and after translating the site
    $\vec{x}_i$ by $\vec{\delta_x}$. Voronoi ge\-ne\-rators are
    represented with dots. The solid lines are the polygons before
    moving $\vec{x_i}$. The dotted lines represent the new polygons
    after varying $\vec{x_i}$.}
  \label{fig:VoronoiDelta}
\end{figure}

To evaluate $\frac{\partial V_k}{\partial x_i}$ we move the generator
$\vec{x}_i$ an infinitesimal quantity $\delta_x$ parallel to the
$\hat{x}$ axis as in Figure \ref{fig:VoronoiDelta}. We will calculate
\begin{equation}
\frac{\partial V_k}{\partial x_i} =
\lim_{\delta_x\to 0}\frac{\Delta V}{\delta_x},\label{eq:DerivadaLimite}
\end{equation}
where $\Delta V_k = V_k(\vec{x}_1, \cdots,
\vec{x}_i + \vec{\delta_x}, \cdots, \vec{x}_{N_V}) -
V_k(\vec{x}_1, \cdots, \vec{x}_i, \cdots, \vec{x}_{N_V})$.

It can be seen in Figure \ref{fig:VoronoiDelta} that when moving the
generator $\vec{x}_i$, the only polygons modified are $\mathcal{V}_i$
and its neighbors. Using this, Eq. \ref{eq:VisDiscreta} leads to
\begin{eqnarray}
\Delta V_k & = & I_i'\int_{\mathcal{V}_i'}f_k(\vec{x})d\vec{x} -
I_i\int_{\mathcal{V}_i}f_k(\vec{x})d\vec{x}\nonumber\\
& & + \sum_{j\in J_i}\bigg(I_j'\int_{\mathcal{V}_j'}f_k(\vec{x})d\vec{x} -
I_j\int_{\mathcal{V}_j}f_k(\vec{x})d\vec{x}\bigg),\label{eq:deltaVis}
\end{eqnarray}
where $\mathcal{V}_i$ is the polygon generated by $\vec{x}_i$ before
moving, $\mathcal{V}_i'$ is the same polygon after moving $\vec{x}_i$,
$J_i$ is the set of indices of the polygons that are neighbors to
$\mathcal{V}_i$ and $J_i'$ is the set of indices of the polygons that
are neighbors to $\mathcal{V}_i'$.

It can be seen in Figure \ref{fig:VoronoiDelta} that
\begin{eqnarray}
\mathcal{V}_i = (\mathcal{V}_i\cap \mathcal{V}_i')\cup(\mathcal{V}_i\setminus \mathcal{V}_i\cap \mathcal{V}_i'), & \mathcal{V}_i' =
(\mathcal{V}_i\cap \mathcal{V}_i')\cup(\mathcal{V}_i'\setminus \mathcal{V}_i\cap \mathcal{V}_i'),\\
\mathcal{V}_j = (\mathcal{V}_j\cap \mathcal{V}_j')\cup(\mathcal{V}_i'\cap \mathcal{V}_j), & \mathcal{V}_j' = (\mathcal{V}_j\cap
\mathcal{V}_j')\cup(\mathcal{V}_i\cap \mathcal{V}_j'),
\end{eqnarray}
so, Eq. \ref{eq:deltaVis} is
\begin{eqnarray}
\Delta V_k & = & (I_i'-I_i)\int_{\mathcal{V}_i'\cap
\mathcal{V}_i}f_k(\vec{x})d\vec{x}\nonumber\\
&& + \sum_{j\in J_i}\bigg[(I_i' - I_j)\int_{\mathcal{V}_i'\cap
\mathcal{V}_j}f_k(\vec{x})d\vec{x} + (I_j' - I_i)\int_{\mathcal{V}_i\cap
\mathcal{V}_j'}f_k(\vec{x})d\vec{x}\bigg].
\end{eqnarray}
In our case the cells' intensities don't depend of the position of the
generators, so we obtain
\begin{equation}\label{eq:DeltaVis}
\Delta V_k = \sum_{j\in J_i}\bigg[(I_i-I_j)\bigg(\int_{\mathcal{V}_i'\cap
\mathcal{V}_j}f_k(\vec{x})d\vec{x} - \int_{\mathcal{V}_i\cap
\mathcal{V}_j'}f_k(\vec{x})d\vec{x}\bigg)\bigg].
\end{equation}

\begin{figure}[h]
  \epsscale{.50}
%  \plotone{figuras/VoronoiCoordenadasB&N.ps}
  \plotone{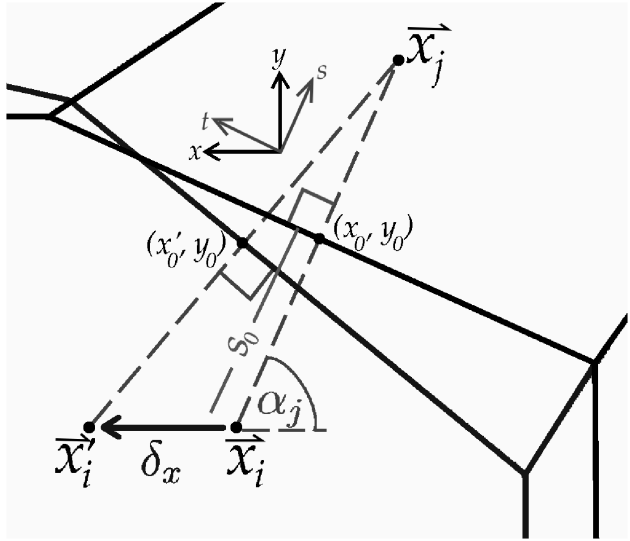}
  \caption{Change of coordinates from $(x, y)$ to $(s, t)$.}
  \label{fig:VoronoiCoordenadas}
\end{figure}

It can be seen in Figure \ref{fig:VoronoiDelta} that to obtain $\Delta
V_k$ we must integrate only over the shaded regions.  For this
purpose, for each region between $\vec{x}_i$ and $\vec{x}_j$ we will
define a coordinate system
\begin{eqnarray}
\hat{s} = -\cos{\alpha_j}\hat{x} + \sin{\alpha_j}\hat{y}. & \hat{t}
= \sin{\alpha_j}\hat{x} + \cos{\alpha_j}\hat{y},\label{eq:sistema_coordenadas}
\end{eqnarray}
where $\alpha_j$ is the angle formed by the $-\hat{x}$ axis and the
edge $a_{ij}$ between $\vec{x}_i$ and $\vec{x}_j$ (see Figure
\ref{fig:VoronoiCoordenadas}). Using this change of coordinates, the
integral over the region of interest is
\begin{equation}
\int_{\mathcal{V}_i'\cap \mathcal{V}_j}f_k(x, y)dxdy = \int_{\mathcal{V}_i'\cap
\mathcal{V}_j}f_k(s,t)dsdt.\label{eq:int}
\end{equation}

Let $\vec{x_i} = (x_i, y_i)$ be the position of the $i$ cell's
generator, $\vec{x_j} = (x_j, y_j)$ one of its neighbor, and
$\vec{x_i}' = (x_i + \delta_x, y_i)$ the site's position after moving
it a quantity $\delta_x$. We define $\vec{x_0} \equiv (x_0, y_0)$ as
the point in the intersection of the segment formed by $\vec{x_i}$ and
$\vec{x_j}$ and its respective edge $a_{ij}$. The same way, we define
$\vec{x_0}' = (x_0', y_0')$ as the point in the intersection of the
segment formed by $\vec{x_i}'$ and $\vec{x_j}$ and its respective edge
$a_{ij}'$. It can be seen on Figure \ref{fig:VoronoiCoordenadas} that
$x_0 = \frac{x_i + x_j}{2}$ , $x_0' = x_0 + \frac{\delta}{2}$ and $y_0'
= y_0 = \frac{y_i + y_j}{2}$.

The edge $a_{ij}$ is defined in the new coordinate system by
\begin{equation}
s = s_0 = -x_0\cos{\alpha_j} + y_0\sin{\alpha_j}.
\end{equation}
In the same way, the edge $a_{ij}'$ is defined in the original
coordinate system by
\begin{equation}
y = m(x - x_0') + y_0,
\end{equation}
where
\begin{equation}
m \equiv \frac{x_i + \delta_x - x_j}{y_j - y_i}.
\end{equation}
We can define the same line in our new coordinate system as
\begin{equation}
s = m't+b',
\end{equation}
where
\begin{eqnarray}
m' & \equiv & -\frac{\cos{\alpha_j} + m\sin{\alpha_j}}{\sin{\alpha_j}
  - m\cos{\alpha_j}},\\ 
b' & \equiv & \frac{-mx_0' +
  y_0}{\sin{\alpha_j} - m\cos{\alpha_j}}.
\end{eqnarray}
This can be approximated to first order in $\delta_x$ as
\begin{eqnarray}
m' & \simeq & \delta_xM_x,\\
b' & \simeq &s_0 + \delta_xB_x,
\end{eqnarray}
where
\begin{eqnarray}
M_x & \equiv &\frac{\sin^2\alpha_j}{y_j - y_i} =
\frac{\sin{\alpha_j}\cos{\alpha_j}}{x_i - x_j},\\ 
B_x & \equiv &
\frac{\sin{\alpha_j}}{y_j-y_i}(s_0\cos{\alpha_j} + x_i) =
\frac{\cos{\alpha_j}}{x_i-x_j}(s_0\cos{\alpha_j} + x_i).
\end{eqnarray}

The integral in Eq. \ref{eq:DeltaVis} using our new coordinate system
will be
\begin{eqnarray}
\mathcal{I} & = & \int_{\mathcal{V}_i'\cap \mathcal{V}_j}f_k(\vec{x})d\vec{x} - \int_{\mathcal{V}_i\cap
\mathcal{V}_j'}f_k(\vec{x})d\vec{x}\\
& = & \int\int_{a_{ij}}^{a_{ij}'}A(\vec{x})e^{2\pi i(ux + vy)}dxdy.\label{eq:Ixy}
\end{eqnarray}
If we use $A(\vec{x})$ in the $(s, t)$ coordinate system as a
pixelated image, Eq. \ref{eq:Ixy} will be
\begin{equation}
\mathcal{I} = \sum_{l\ \epsilon \mathrm{\ pixeles\ de\ }
a_i}A_l\int_{t_{ijl}^1}^{t_{ijl}^2}\int_{s_0}^{m't+b'} e^{2\pi i(ux(s,
t) + vy(s, t))}dsdt,
\end{equation}
where $t_{ijl}^1$ and $t_{ijl}^2$ are the $t$ coordinate of the
beginning and end of the portion of the edge $a_{ij}$ that intersects pixel
$l$.  Developing the previous expression,
\begin{eqnarray}
\mathcal{I} & = & \sum_{l}A_l\int_{t_{ijl}^1}^{t_{ijl}^2}\int_{s_0}^{m't+b'}
e^{2\pi i(u(-s\cos{\alpha_j} + t \sin{\alpha_j}) + v(s\sin{\alpha_j} +
t \cos{\alpha_j}))}dsdt,\label{eq:integralPixel}\\ 
% & = &
% \sum_{l}A_l\int_{t_{ijl}^1}^{t_{ijl}^2}\int_{s_0}^{m't+b'} e^{2\pi
% i(sc_1 + tc_2)}dsdt,\\ 
& \simeq & 
\sum_{l}\frac{A_l}{\pi c_2}e^{2\pi i(s_0c_1 + \bar{t}_{ijl}c_2)}\kappa_{ijl}\delta_x,
\label{eq:IExacto}
\end{eqnarray}
where we defined
\begin{eqnarray}
  c_1 & \equiv & -u\cos{\alpha_j} + v\sin{\alpha_j},\\ 
  c_2 & \equiv & u\sin{\alpha_j} + v\cos{\alpha_j},\\ 
  \kappa_{ijl} & \equiv &
  (M_x\bar{t}_{ijl} + B_x)\sin{(\pi c_2\Delta t_{ijl})} \\\nonumber 
  &&+ i\frac{M_x}{2} \bigg(\frac{\sin(\pi c_2\Delta t_{ijl})}{\pi c_2} -
  \Delta t_{ijl}\cos{(\pi c_2\Delta t_{ijl})}\bigg),\\ 
  \bar{t}_{ijl} & \equiv & \frac{t_{ijl}^1 + t_{ijl}^2}{2},\\ 
  \Delta t_{ijl} & \equiv & \frac{t_{ijl}^2 - t_{ijl}^1}{2}.
\end{eqnarray}

In the calculation above we integrated over the fraction of the edge
that falls inside pixel $l$ and then summed these integrals over the
whole edge $a_i$. It is also possible to approximate the integral of
Eq. \ref{eq:integralPixel} as $\int_{t_{ijl}^1}^{t_{ijl}^2}g(t) dt =
g(\bar{t}_{ijl})\Delta t_{ijl}$, which is equivalent to taking the
limit over the integral $\mathcal{I}$ of Eq. \ref{eq:IExacto},
$\lim_{\Delta t_{ijl}\rightarrow 0} \mathcal{I}$, obtaining
\begin{equation}
  \mathcal{I} = \sum_{l} A_l\Delta t_{ijl} (M_x\bar{t}_{ijl} + B_x)e^{2\pi
    i(\bar{t}_{ijl}c_2 + s_0c_1)}\delta_x.\label{eq:IAprox}
\end{equation}

We found by direct evaluation that the difference between
Eq.~\ref{eq:IAprox} and Eq.~\ref{eq:IExacto} is negligible, so, for
simplicity, we will use Eq. \ref{eq:IAprox}. Introducing
Eq. \ref{eq:IAprox} in Eq. \ref{eq:DeltaVis}, we obtain
\begin{equation}
  \Delta V_k = \delta_x\sum_{j\in
    J_i}\bigg[(I_i-I_j)\sum_{l} A_l\Delta t_{ijl} (M_x\bar{t}_{ijl} +
    B_x)e^{2\pi i(\bar{t}_{ijl}c_2 + s_0c_1)}\bigg],
\end{equation}
so, according to Eq. \ref{eq:DerivadaLimite}, the derivative of the
$k$ visibility with respect to the position $x$ of polygon $i$ is
\begin{eqnarray}
  \frac{\partial V_k}{\partial x_i} & = &
  \lim_{\delta_x\to 0}\frac{\Delta V}{\delta_x},\\
  & = & \sum_{j\in
    J_i}\bigg[(I_i-I_j)\sum_{l} A_l\Delta t_{ijl} (M_x\bar{t}_{ijl} +
    B_x)e^{2\pi i(\bar{t}_{ijl}c_2 + s_0c_1)}\bigg].
\end{eqnarray}
Similarly, for the derivative with respect to the position $y$ of the
$i$ polygon we obtain
\begin{equation}
  \frac{\partial V_k}{\partial y_i} = \sum_{j\in
    J_i}\bigg[(I_i-I_j)\sum_{l} A_l\Delta t_{ijl} (M_y\bar{t}_{ijl} +
    B_y)e^{2\pi i(\bar{t}_{ijl}c_2 + s_0c_1)}\bigg],
\end{equation}
where
\begin{eqnarray}
M_y & \equiv &\frac{\cos^2\alpha_j}{x_i - x_j} =
\frac{\sin{\alpha_j}\cos{\alpha_j}}{y_j - y_i},\\ 
B_y & \equiv & \frac{\sin{\alpha_j}}{y_j-y_i}(s_0\sin{\alpha_j} - y_i)
= \frac{\cos{\alpha_j}}{x_i-x_j}(s_0\sin{\alpha_j} - y_i).
\end{eqnarray}

\subsection {Calculation of the Derivatives of $S$}\label{ap:derivativesdS}

We defined our entropy as 
\begin{eqnarray}
S & = & \ln\left(\frac{N!}{n^N
  \prod_{i=1}^{n}N_{i}!}\right) \\
  & = & \ln(N!) - N\ln(n) - \sum_{i=1}^{n}\ln(N_i!)\\
  & = & \ln\Big(\Gamma(N + 1)\Big) - N\ln(n) - \sum_{i=1}^{n}\ln\Big(\Gamma(N_i +
  1)\Big),
\end{eqnarray}
where $N_i = \frac{I_i}{\sigma_\mathrm{q}}$ is the number of quanta
in cell $i$, $N = \sum_iN_i$ and $\Gamma$ is the Gamma function. It
can be seen that this function does not depend on the position of the
Voronoi generators, so
\begin{equation}
  \frac{\partial S}{\partial x_i} = \frac{\partial S}{\partial y_i} = 0.
\end{equation}

Using Weierstrass' definition of the Gamma function
\[\Gamma(z) = z^{-1}e^{-\gamma z}\prod_{n=1}^\infty
  \left[\left(1 + \frac{z}{n}\right)^{-1} e^{z/n}\right],\]
where $\gamma$ is Euler's constant, we can obtain
\begin{equation}
  \frac{\partial \ln \Big(\Gamma(z + 1)\Big)}{\partial z} = -\gamma +
  \sum_{n=1}^z\frac{1}{n}
\end{equation}
so, the derivative of $S$ with respect to $I_i$ is
\begin{equation}
  \frac{\partial S}{\partial I_i} = \frac{1}{\sigma_\mathrm{q}}
  (\sum_{k=1}^{N}\frac{1}{k} - \ln{n} - \sum_{k=1}^{N_i}\frac{1}{k}).
\end{equation}

\subsection {Finite Difference Cross Check on the Derivatives}

Numerical calculation of the derivatives by finite differences is not
very accurate, in particular for the position of the
generators. Finite difference derivatives are calculated as
$\frac{\partial L}{\partial x} = \frac{ L(x + \delta) -
L(x)}{\delta}$, where $\delta$ is a small displacement of $x$. In the
case of the positions of the generators, if $\delta$ is too small, the
pixelization of the Voronoi diagram (needed to obtain the model
visibilities) will not change after the displacement $\delta$. On the
other hand, if $\delta$ is too big, the generator displacement may
cause the function to change abruptly, as explained below. That is why
we calculated the analytical expression for the derivatives.

\begin{figure}[h]
  \epsscale{1.1}
%  \plottwo{figuras/dx.ps}{figuras/dI.ps}
  \plottwo{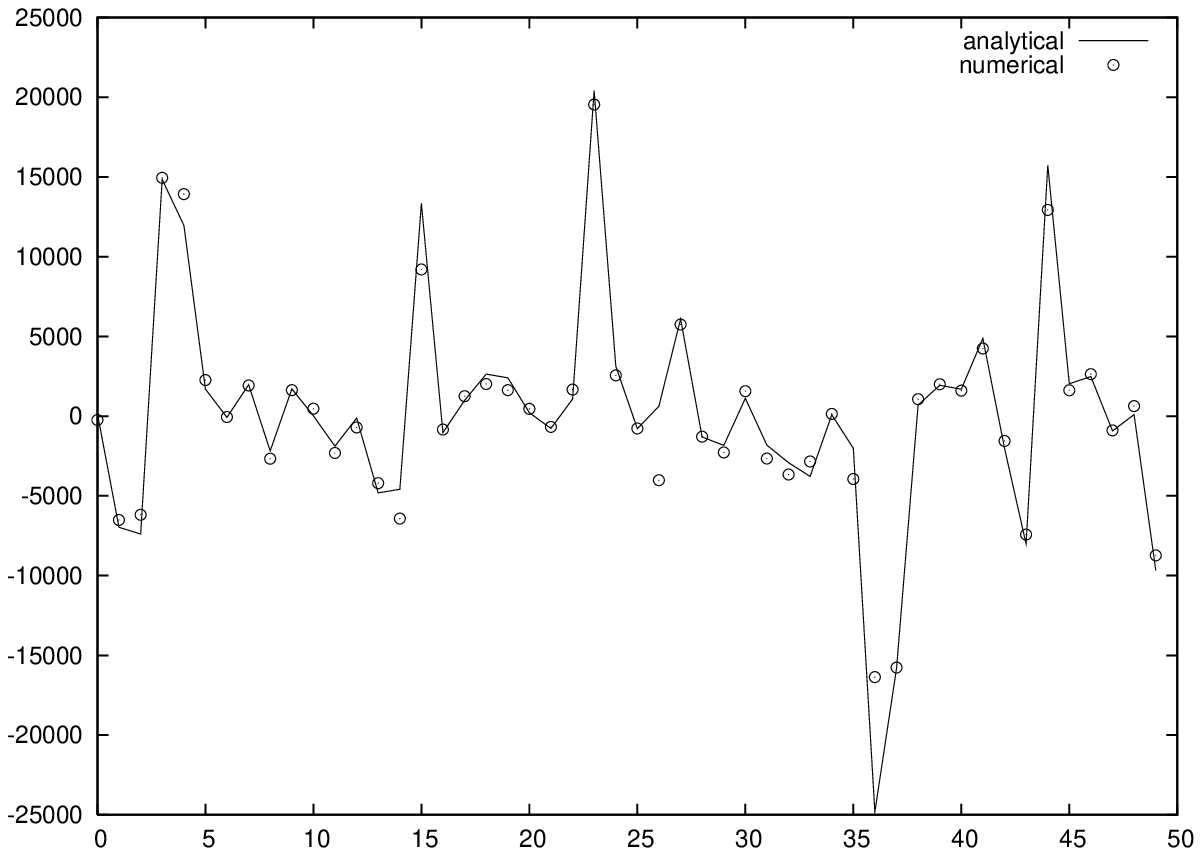}{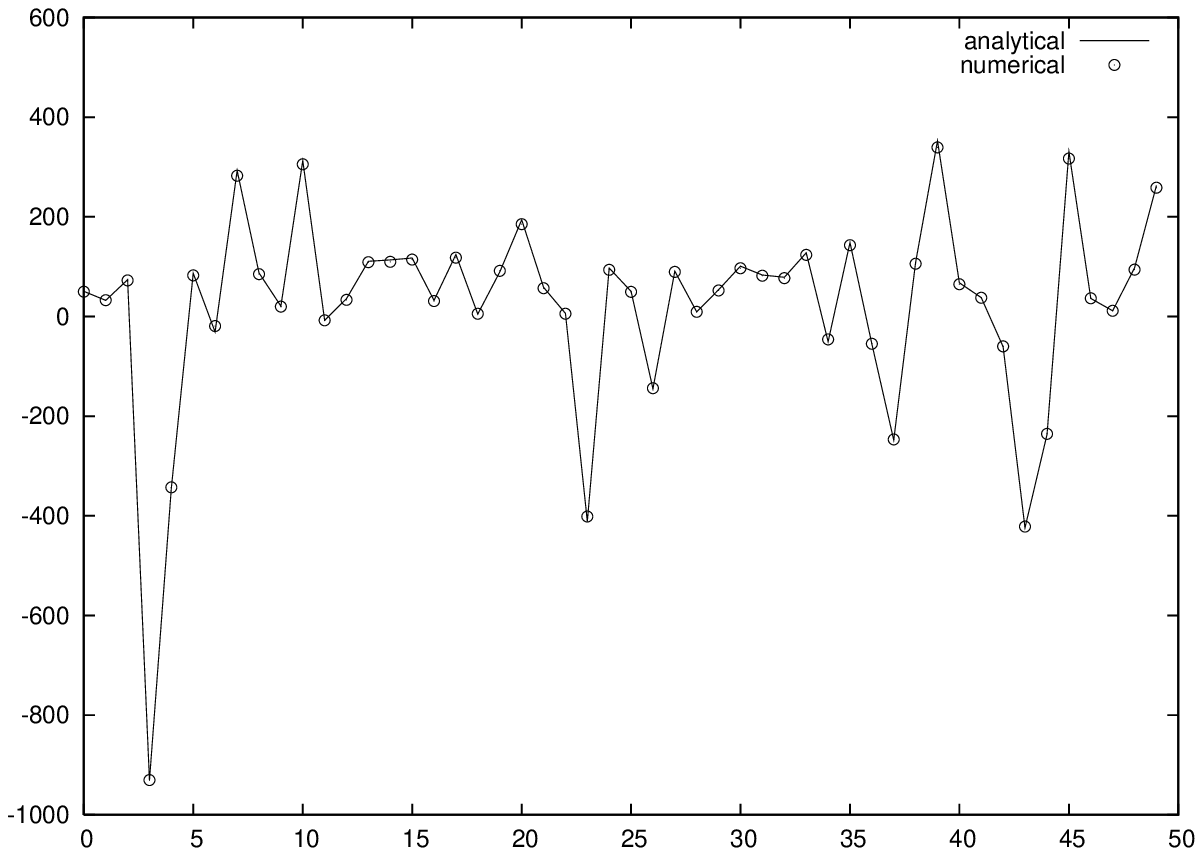}
  \caption{Verification of the derivatives. The solid line shows
  analytical derivatives, and dots show numerical
  approximations. \emph{Left:} $\frac{\partial L}{\partial
  x_i}$. \emph{Right:} $\frac{\partial L}{\partial I_i}$. 
  The polygon identifier $i$ is indicated on the $x$-axis.
  % The polygon $i$ is represented on the $x$-axis.
  \label{fig:dLdx}}
\end{figure}

To verify that our derivatives are correctly calculated and
programmed, we compared our analytical result with a numerical
calculation. We created a Voronoi tessellation of $50$ polygons with
random positions and intensities and calculated the analytical and
numerical derivatives using these parameters $\{x_i, y_i, I_i\}$. For
the case of $\frac{\partial L}{\partial x_i}$ and $\frac{\partial
L}{\partial y_i}$ this numerical cross check consists of moving each
Voronoi generator a quantity $\delta$ from -0.1 to 0.1 with an
interval of $10^{-3}$ in units of the total size of the square
image. We evaluate the merit function $L$ at each position intervals,
thus obtaining two sequences $\{L_i\}_{i=1}^{2\times 10^2}$. We then
fitted a polynomial of order four to the curve defined by each
sequence $\{L_i\}$ and calculated the derivative of the polynomial at
$\delta = 0$. For the case of $\frac{\partial L}{\partial I_i}$ we
varied the intensity of cell $i$ from $-\sigma_\mathrm{q}$ to
$\sigma_\mathrm{q}$ and did the same approximation to a polynomial of
order four and calculated its derivative. Figure~\ref{fig:dLdx} shows
this cross check for $\frac{\partial L}{\partial x_i}$ and
$\frac{\partial L}{\partial I_i}$. Although the derivatives are
similar, they are not exactly the same for $\frac{\partial L}{\partial
x_i}$. This is caused by the polynomial coarseness fit, as explained
below.

\begin{figure}[h!]
  \epsscale{0.4}
  \plotone{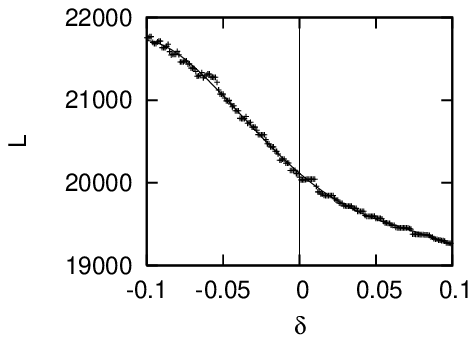}

  \plotone{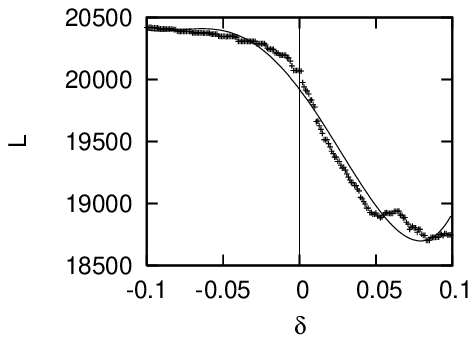}

  \plotone{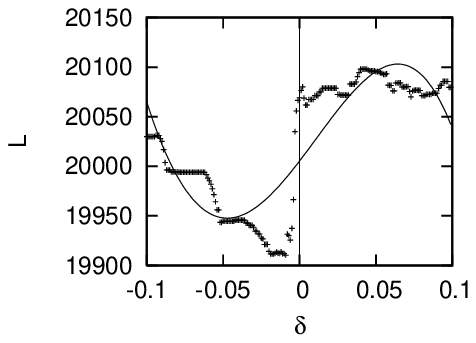}
  \caption{Examples of polynomial fits, used to determine numerical
  derivatives of the optimization function $L$ for a particular
  generator. Dots represent $L$ vs the polygon displacement $\delta$
  in $x$ and the solid line shows the fourth order polynomial fit to
  $L$. A vertical line is drawn at $\delta = 0$, where the derivatives
  were calculated. \emph{Top:} Generator number 37, with a
  satisfactory polynomial fit. \emph{Middle:} Generator number 36, the
  curve is not a good fit at $\delta = 0$. \emph{Bottom:}
  $18^\mathrm{th}$ generator, the curve is not a good fit because $L$
  shows an abrupt variation near $\delta = 0$, which is due to the
  proximity of another generator.\label{fig:ajuste}}
  
  % does not fit well to $L$ because the generator is too close to another
  % one.\label{fig:ajuste}}
\end{figure}

\begin{figure}[h]
  \epsscale{0.9}
%  \plottwo{figuras/VorDelta1.eps}{figuras/VorDelta2.eps}
  \plottwo{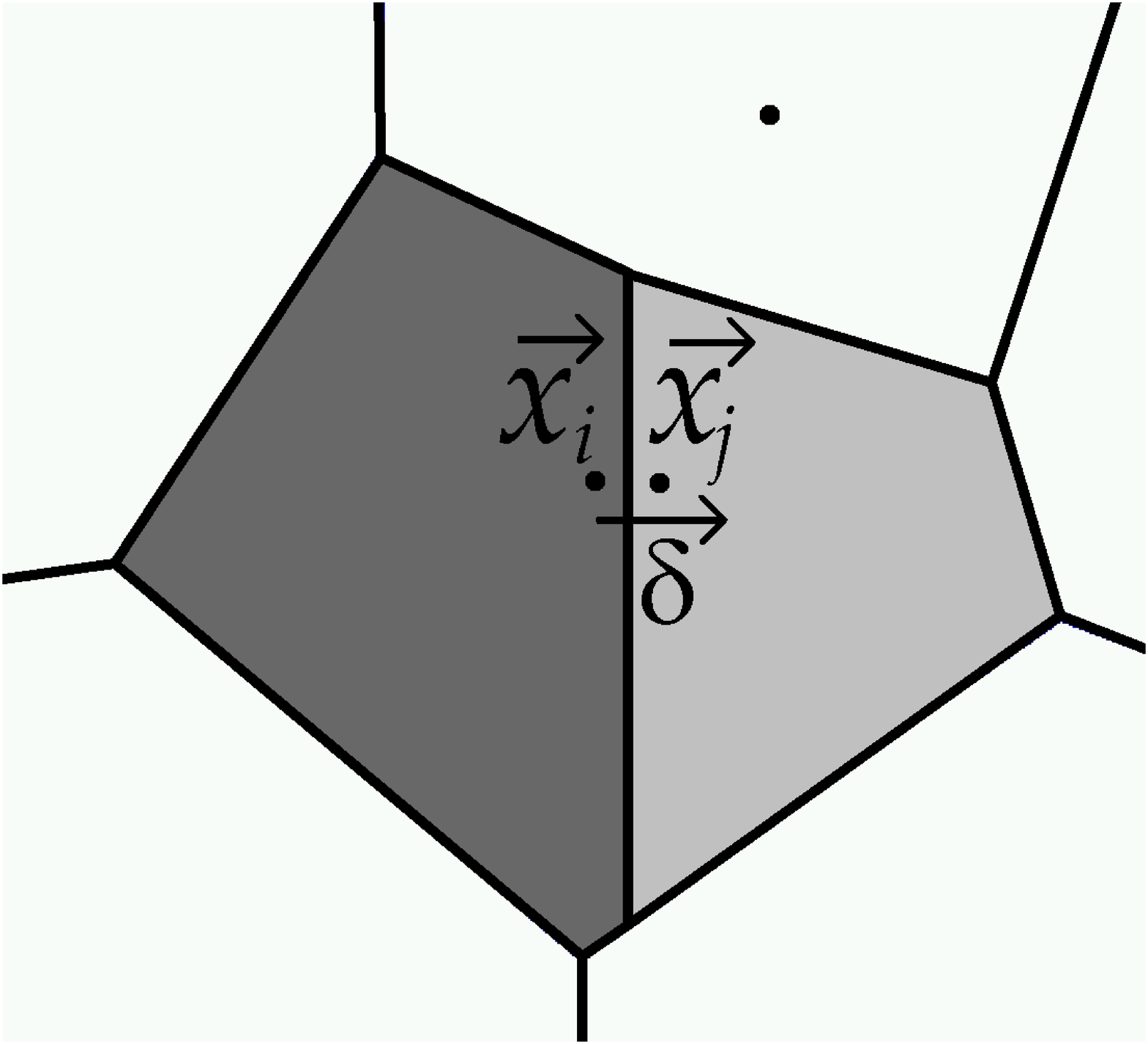}{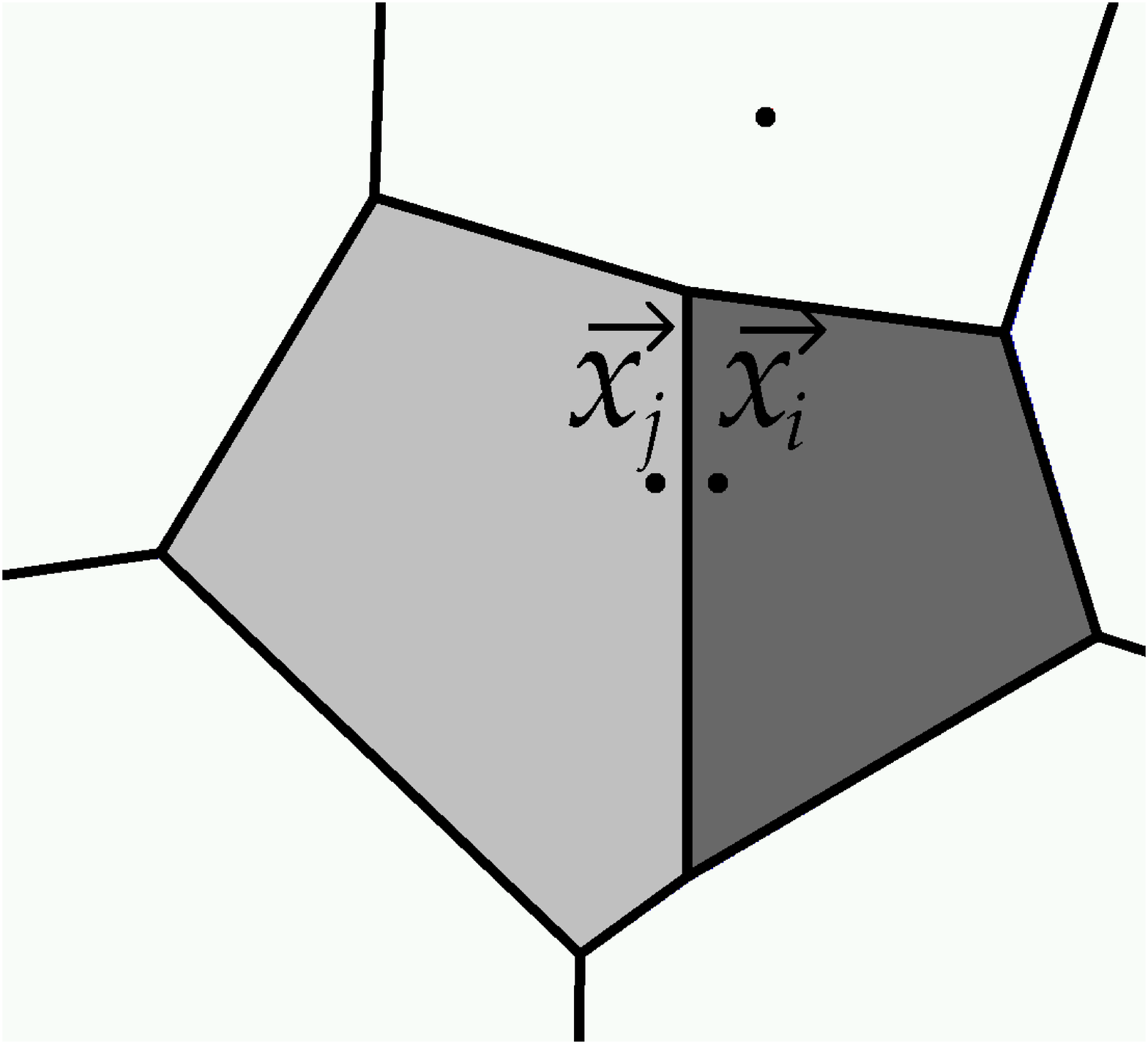}
  \caption{Translation of a generator close to another. \emph{Left:}
  Before moving generator $\vec{x_i}$, polygon $i$, the darker polygon
  in the image, is on the left. \emph{Right:} After moving generator
  $\vec{x_i}$, by a displacement of $\delta$, polygon $i$ is on the
  right of polygon $j$.  When displacing generator $\vec{x_i}$ the
  diagram changes considerably, with a concomitant abrupt variation in
  $L$.
  %making the optimization function $L$ to change as well.
  \label{fig:VorProblem}}
\end{figure}

Figure \ref{fig:ajuste} shows the curve fit for $\frac{\partial
L}{\partial x_i}$ for three different generators (generator number 37,
36 and 18 respectively). It can be seen in Figure \ref{fig:ajuste}
that the polynomial fit adjusts quite well to the function values for
polygon number 37, so on Figure \ref{fig:dLdx} both derivatives are
the same. On the contrary, for polygons number 36 and 18, the fitted
polynomial does not resemble the function $L$ at $\delta = 0$, causing
a slight difference in their derivatives on Figure \ref{fig:dLdx}. For
polygon number 18 the polynomial does not fit the curve at all. This
is the main problem of using a numerical approximation for the
derivatives of $\{\vec{x_i}\}$: when two polygons are closer than
$\delta$, the generator displacement causes the function $L$ to change
abruptly (see Figure \ref{fig:VorProblem}).

 It can be seen that the analytical and numerical derivatives on
Figure \ref{fig:dLdx} are almost the same. As explained above,
differences are produced because there are cases where the polynomials
do not fit well to the variations of the merit function $L$ (for
example, when two generators are too close). In an accurate
calculation it is necessary to use the analytical derivatives.

\section{Fitting a Voronoi Tessellation to an Image}\label{ap:fitting}
Once we have a reasonable reconstruction for a pixelated image, we
would like to fit a Voronoi tessellation to it in order to have a good
initial starting point for the CG. This is done in an incremental way.

We start with a mesh consisting in only one polygon. We calculate the
error per polygon as
\begin{equation}
  e_i^2 = \sum_l
  (I_i - I^\mathrm{im}_l)^2,
\end{equation}
where the sum runs over all the pixels that fall inside polygon $i$,
$I_i$ is the intensity of that polygon and $I^\mathrm{im}_l$ is the
intensity of pixel $l$ in the image to be fitted. In each iteration we
add a new polygon inside the one with the greatest error. The new
generator is inserted in the position of the pixel that has the most
different intensity value with respect to the mesh intensity.

% I'm not sure where to put this. I guess its an incremental algorithm,
% so I left it here.

% {\it Facilities:} \facility{CBI}.

\clearpage

\clearpage

\end{document}